\documentclass[aps,preprint,prd,showpacs,showkeys,nofootinbib,tightenlines]{revtex4}

\usepackage{graphicx}  
\usepackage{amsmath}
\usepackage{bm}  
\usepackage{ulem}
\usepackage{amssymb}
\usepackage{comment}

\newcommand{\bea}{\begin{eqnarray}}
\newcommand{\eea}{\end{eqnarray}}
\newcommand{\beq}{\begin{equation}}
\newcommand{\eeq}{\end{equation}}
\newcommand{\bqa}{\begin{eqnarray}}
\newcommand{\eqa}{\end{eqnarray}}

\begin{document}

\title{
Estimates of the $\bm{X(3872)}$ Cross Section\\ at a Hadron Collider}

\author{Eric Braaten}
\email{braaten.1@osu.edu}
\affiliation{Department of Physics,
         The Ohio State University, Columbus, OH\ 43210, USA}

\author{Li-Ping He}
\email{he.1011@buckeyemail.osu.edu}
\affiliation{Department of Physics,
         The Ohio State University, Columbus, OH\ 43210, USA}

\author{Kevin Ingles}
\email{ingles.27@buckeyemail.osu.edu}
\affiliation{Department of Physics,
         The Ohio State University, Columbus, OH\ 43210, USA}

\date{\today}

\begin{abstract}
The claim that the $X(3872)$ meson cannot be a charm-meson molecule
because its prompt production cross section at hadron colliders is too large
is based on an upper bound  in terms of a cross section for producing charm-meson pairs.
Assuming  $X$ is sufficiently weakly bound, we derive an equality between the $X$ cross section 
and a charm-meson pair cross section that takes into account the threshold enhancement from the $X$ resonance.
The cross section for producing  $X$ is equal to that for producing $D^{*0} \bar{D}^0$ integrated up to a
relative momentum $k_\mathrm{max} = 7.7\, \gamma_X$, where $\gamma_X$ is the binding momentum of $X$.
We also derive an order-of-magnitude estimate of the $X$ cross section in terms of
 a naive charm-meson pair cross section that does not take into account the threshold enhancement, 
such as that produced by a Monte Carlo event generator.
The cross section for producing  $X$ can be approximated by  the naive cross section for producing $D^{*0} \bar{D}^0$ integrated up to a relative momentum $k_\mathrm{max}$ of order  $(m_\pi^2 \gamma_X)^{1/3}$.
The estimates of the prompt $X$ cross section at hadron colliders are consistent 
with the cross sections observed at the Tevatron and the LHC.
\end{abstract}

\smallskip
\maketitle


\section{Introduction}
\label{sec:Introduction}

The discovery of a large number of exotic hadrons containing a heavy quark and its antiquark 
presents a major challenge to our understanding of QCD
\cite{Chen:2016qju,Hosaka:2016pey,Lebed:2016hpi,Esposito:2016noz,Guo:2017jvc,Ali:2017jda,Olsen:2017bmm,Karliner:2017qhf,Yuan:2018inv,Brambilla:2019esw}.
The first of these exotic hadrons to be discovered was the $X(3872)$ meson.
It was discovered in 2003 in exclusive decays of $B^\pm$ mesons into $K^\pm X$ 
through the decay of $X$ into $J/\psi\, \pi^+\pi^-$ \cite{Choi:2003ue}.
Its existence was quickly verified through inclusive production in $p \bar p$ collisions \cite{Acosta:2003zx}.
The $J^{PC}$ quantum numbers of $X$ were eventually determined to be $1^{++}$ \cite{Aaij:2013zoa}.
Its mass  is extremely close to the $D^{*0} \bar D^0$  threshold,
with the difference being only $0.01 \pm 0.18$~MeV \cite{Tanabashi:2018oca}.
This suggests that $X$ is a weakly bound S-wave charm-meson molecule
with the flavor structure
\begin{equation}
\big| X(3872) \rangle = \frac{1}{\sqrt2}
\Big( \big| D^{*0} \bar D^0 \big\rangle +  \big| D^0 \bar D^{*0}  \big\rangle \Big).
\label{Xflavor}
\end{equation}

The $X$ can be produced by any reaction that can produce its constituents $D^{*0} \bar D^0$ and $D^0 \bar D^{*0}$.
In particular, it can be produced in high energy hadron collisions.
The inclusive production of $X$ in $p \bar p$ collisions has been studied at the Tevatron by the 
CDF \cite{Acosta:2003zx} and D0 \cite{Abazov:2004kp} collaborations.
The inclusive production of $X$ in $p p$ collisions has been studied at the 
Large Hadron Collider (LHC) by the LHCb \cite{Aaij:2011sn},
CMS \cite{Chatrchyan:2013cld}, and ATLAS \cite{Aaboud:2016vzw} collaborations.
At a high energy hadron collider,  $X$ is produced by the weak decays of bottom hadrons
and by QCD mechanisms that create charm quarks and antiquarks.
If $X$ is produced by the weak decays of bottom hadrons, its decay products emerge from a vertex 
displaced from the collision point. If $X$ is produced by QCD mechanisms, 
its decay products emerge from the collision point, so these mechanisms are referred to as {\it prompt} production.
Cross sections for inclusive prompt  production of $X$ have been measured by the 
CDF \cite{Acosta:2003zx}, CMS \cite{Chatrchyan:2013cld}, and ATLAS \cite{Aaboud:2016vzw} collaborations.

The substantial prompt production rate of $X$ at hadron colliders has often been used as an
argument against its identification as a charm-meson molecule.  
This argument is based on an upper bound on the cross section for producing $X$
in terms of the cross section for producing the charm-meson pair $D^{*0} \bar D^0$
integrated up to a maximum relative momentum $k_\mathrm{max}$ \cite{Bignamini:2009sk}.
The estimate for $k_\mathrm{max}$  in Ref.~\cite{Bignamini:2009sk} was 
approximately the binding momentum $\gamma_X$ of the $X$.
In Ref.~\cite{Artoisenet:2009wk}, it was pointed out that the derivation of the upper bound 
in Ref.~\cite{Bignamini:2009sk} requires $k_\mathrm{max}$ to be of order the pion mass $m_\pi$ instead of $\gamma_X$.
In this paper, we use the methods of  Ref.~\cite{Artoisenet:2009wk} 
to derive equalities between the $X$ cross section and 
$D^{*0} \bar D^0$ cross sections integrated up to  $k_\mathrm{max}$.  If we take into account the 
threshold enhancement due to the $X$ resonance, the value of $k_\mathrm{max}$ is $7.7\gamma_X$. 
If we use a naive $D^{*0} \bar D^0$ cross section without the threshold enhancement,
the value of $k_\mathrm{max}$ is order $(m_\pi^2 \gamma_X)^{1/3}$.
The resulting estimates of the prompt cross sections are compatible with the measurements at the Tevatron and the LHC.

The outline of this paper is as follows.
In Section~\ref{sec:Molecule}, we describe some universal aspects of weakly bound S-wave molecules
and the scattering of their constituents.
In  Section~\ref{sec:CrossSection}, we present experimental upper and lower bounds on the cross sections 
for the production of $X$ at the Tevatron and the LHC.
In  Section~\ref{sec:Upperbound}, we discuss the theoretical upper bound on the prompt cross section for producing  $X$
at hadron colliders derived in Ref.~\cite{Bignamini:2009sk}.
 In  Section~\ref{sec:Equality}, we derive equalities between the $X$ cross section
and a charm-meson pair cross section with and without the threshold enhancement from the $X$ resonance.
In Section~\ref{sec:Summary}, we summarize our results and discuss their implications.


\section{ Bound S-wave  Molecule}
\label{sec:Molecule}

If short-range interactions produce an S-wave bound state extremely close to a scattering threshold,
the few-body physics has universal aspects that are determined by the {\it binding momentum} $\gamma_X$ 
of the bound state \cite{Braaten:2004rn}. 
The binding energy is $\gamma_X^2/2\mu$, where $\mu$ is the reduced mass of the constituents.
The momentum-space wavefunction in the region of the relative momentum $k$ 
below the inverse range has the universal form
\begin{equation}
\psi_X(k) = \frac{ \sqrt{8 \pi \gamma_X}}{k^2 + \gamma_X^2}.
\label{psiX-k}
\end{equation}
The low-energy scattering of the constituents also has universal  aspects determined by $\gamma_X$
through a simple function of the complex energy $E$ relative to the scattering threshold:
\begin{equation}
f_X(E) = \frac{1}{-\gamma_X+\sqrt{-2\mu E }}.
\label{f-E}
\end{equation}
This function has a branch cut along the positive $E$ axis and a pole at $E=-\gamma_X^2/2 \mu$.
The universal elastic scattering amplitude  in the region of relative momentum $k$ below the inverse range 
is obtained by evaluating this function at energy $E=k^2/2\mu + i \epsilon$.

The analytic function $f_X(E)$ also gives the energy distribution from creation of the constituents at short distances. 
 By the optical theorem, the distribution in the energy $E$ 
below the energy scale set by  the range  is proportional to the imaginary part of $f_X(E)$: 
\begin{equation}
\mathrm{Im}[f_X(E+ i \epsilon)] = \frac{\pi \gamma_X}{\mu} \delta(E + \gamma_X^2/2\mu)
+ \frac{\sqrt{2\mu E}}{\gamma_X^2+2\mu E} \theta(E).
\label{Imf-E}
\end{equation}
There is a delta-function term at a negative energy  from the production of the weakly bound molecule
and a theta-function term  with positive energy from the  production of the constituents of the molecule.

\begin{figure}[ht]
\includegraphics*[width=0.8\linewidth]{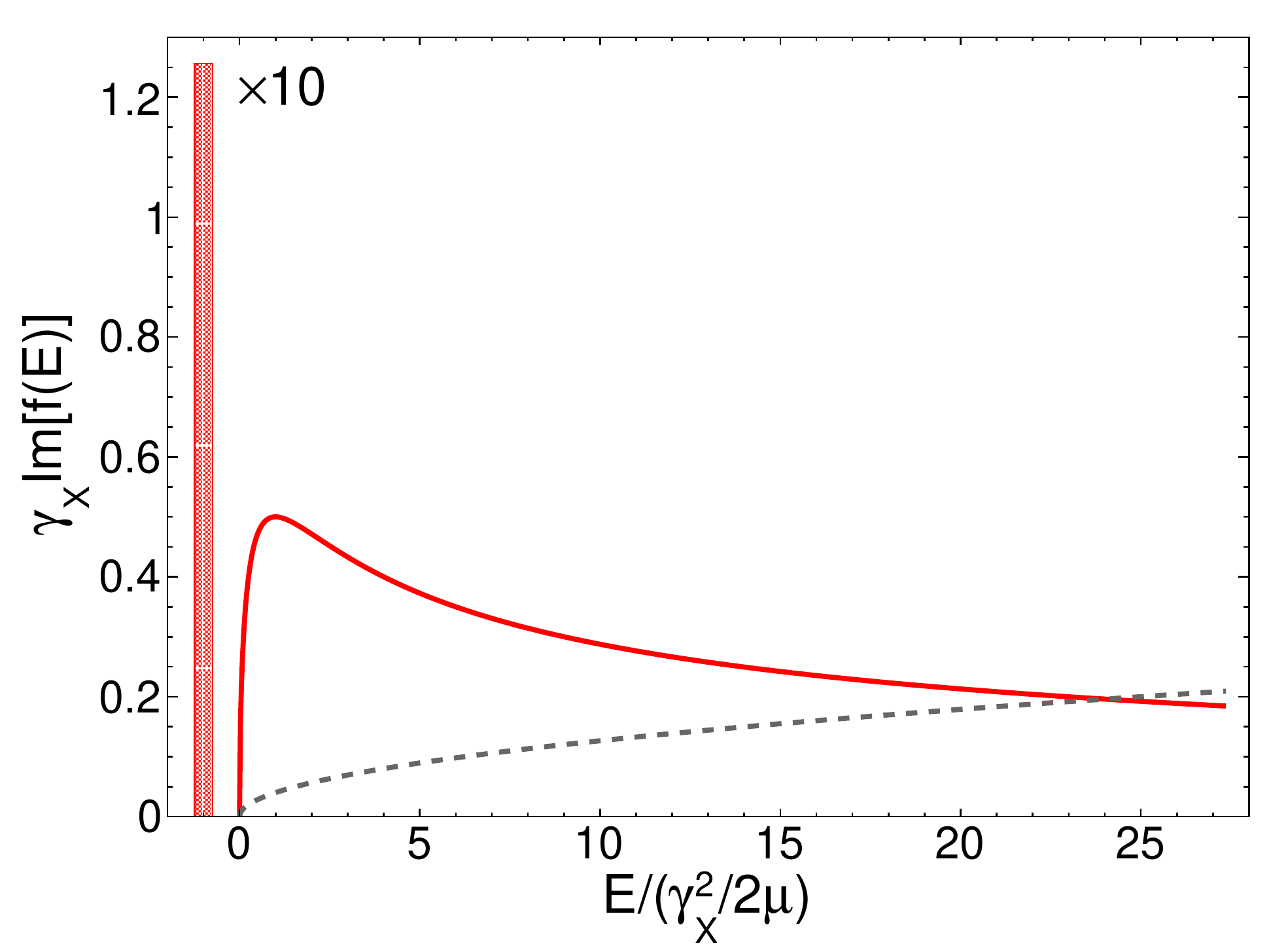} 
\caption{
Universal energy distribution 
$\mathrm{Im}[f_X(E+ i \epsilon)]$ in Eq.~\eqref{Imf-E} as a function of the energy $E$.
The tall rectangle below the scattering threshold at $E=0$ 
represents the delta function from production of the bound state and should be multiplied by 10.
The dashed curve is the naive energy distribution in the absence of the resonance 
given by Eq.~\eqref{Imf-Enaive} with $\Lambda = 5 \, \gamma_X$.
}
\label{fig:DstarDbar}
\end{figure}

The universal energy distribution in Eq.~\eqref{Imf-E} from the creation of the constituents at short distances 
is illustrated in  Fig.~\ref{fig:DstarDbar}.
The delta-function term below the scattering threshold at $E=0$ is represented by a tall rectangle
centered at $E = -\gamma_X^2/2\mu$ which would have the correct area if it was multiplied by 10.
That area is equal to the area of the energy distribution above the threshold 
integrated up to the energy $20.2 \,(\gamma_X^2/2\mu)$,
which corresponds to a relative momentum $k_\mathrm{max} = 4.49\, \gamma_X$.
The energy distribution above the threshold has a maximum at the binding energy $\gamma_X^2/2\mu$.
For $E$ much larger than $\gamma_X^2/2\mu$, the universal energy distribution in Eq.~\eqref{Imf-E}
decreases as $E^{-1/2}$.  This scaling behavior should continue up to the energy scale set by the range.

The naive energy distribution in the absence of the resonance has a form at low energy that can be obtained 
from Eq.~\eqref{Imf-E} by omitting the delta-function term and by replacing the denominator $\gamma_X^2+k^2$
in the theta-function term by $\Lambda^2$, where $\Lambda$ is some momentum 
of order the inverse range:
\begin{equation}
\mathrm{Im}[f_\mathrm{naive}(E+ i \epsilon)] =
 \frac{1}{\Lambda^2}\sqrt{2\mu E} ~\theta(E).
\label{Imf-Enaive}
\end{equation}
The naive energy distribution is shown as a dashed line in Fig.~\ref{fig:DstarDbar}.
The actual energy distribution should cross over from the solid line to the dashed line, 
and it should therefore have a local minimum near $\Lambda^2/2\mu$.
The area under the delta function from the bound state is equal to the area of the naive energy distribution  
 integrated up to the energy $(3 \pi\gamma_X \Lambda^2)^{2/3}/2\mu$,
which corresponds to a relative momentum $k_\mathrm{max} = (3 \pi\gamma_X \Lambda^2)^{1/3}$.

If the $X(3872)$ is a weakly bound charm-meson molecule, its constituents are
the superposition of charm mesons in Eq.~\eqref{Xflavor}.
We denote the masses of the charm mesons $D^0$ and $D^{*0}$ by $M_0$ and $M_{*0}$, 
respectively.
The reduced mass of $D^{*0} \bar D^0$ is $\mu=M_{*0}M_0/(M_{*0}+M_0)$.
The range of the interactions between the charm mesons is $1/m_\pi$,
where $m_\pi$ is the pion mass.
The corresponding energy scale $m_\pi^2/2\mu$  is about 10~MeV.
This is comparable to the energy of the $D^{*+} D^-$ scattering threshold,
which is 8.2~MeV above the  $D^{*0} \bar D^0$  scattering threshold.
The present value of the difference $E_X$ between the mass of the $X$ 
and the energy of the $D^{*0} \bar D^0$ scattering threshold is \cite{Tanabashi:2018oca}
\begin{equation}
E_X \equiv M_X - (M_{*0}+M_0) =( +0.01 \pm 0.18)~\mathrm{MeV}.
\label{eq:deltaMX}
\end{equation}
The central value in Eq.~\eqref{eq:deltaMX} corresponds to a charm-meson pair above the scattering threshold.
The value lower by $1\sigma$ corresponds to a bound state with binding energy $|E_X| =0.17$~MeV
and binding momentum $\gamma_X=18$~MeV.
The upper bound on the binding energy with 90\% confidence level is $|E_X| < 0.22$~MeV.

Some qualitative aspects of the energy distribution illustrated in Fig.~\ref{fig:DstarDbar} have been observed 
by the Belle collaboration in the decays of $B$ mesons into $K D^0 \bar D^0 \pi^0$ \cite{Gokhroo:2006bt}.
The $D^0 \bar D^0 \pi^0$  invariant-mass distribution has a peak near the $D^{*0} \bar D^0$  threshold.
The energy resolution was not sharp enough to resolve the contributions from the narrow peak below
the  $D^{*0} \bar D^0$  threshold from the $X$ bound state and the peak above the threshold
from the $D^{*0} \bar D^0$  and $D^{*0} \bar D^{*0}$ threshold enhancements.
The difference between the fitted curve to the 
$D^0 \bar D^0 \pi^0$ invariant-mass distribution and the combinatorial background in Ref.~\cite{Gokhroo:2006bt}
has a minimum at a $D^0 \bar D^0 \pi^0$ energy about 11~MeV above the $D^{*0} \bar D^0$ threshold.
If we set $\Lambda^2/2 \mu = 11$~MeV, we get an estimate for $\Lambda$ of about 150~MeV.
This is consistent with $\Lambda$ being of order $m_\pi$.


\section{Cross sections for $\bm{X}$ at Hadron Colliders}
\label{sec:CrossSection}

In this section, we summarize experimental results on the inclusive prompt cross sections for $X(3872)$
at the Tevatron and the LHC. We use them to obtain upper and lower bounds on the cross sections.

Within months of the discovery of the $X$ in $B$ meson decays by the Belle Collaboration 
in September 2003 \cite{Choi:2003ue},
its existence was confirmed by the CDF Collaboration through inclusive production of $X$ in $p \bar p$ collisions 
at the Tevatron \cite{Acosta:2003zx}.  The $X$ was observed in the discovery decay mode $J/\psi\, \pi^+\pi^-$.
Some of the $X$ events were produced from decays of bottom hadrons, 
with the ultimate decay products $\mu^+\mu^- \pi^+\pi^-$ emerging from a displaced vertex.
The majority of $X$ events were produced promptly, with the ultimate decay products 
emerging from the primary collision vertex.  
The prompt $X$ events presumably come from QCD production mechanisms.
The CDF Collaboration has reported ratios of
the prompt cross sections for $X$ and $\psi(2S)$ modulo the ratio $\epsilon_\psi/\epsilon_X$
of the efficiencies for observations of $\psi(2S)$ and $X$ in their $J/\psi\, \pi^+\pi^-$ decay modes \cite{Bauer:2005yu}.
The ratio $\epsilon_\psi/\epsilon_X$ is presumably closer to 1 than to 2 or 1/2.
In Ref.~\cite{Bignamini:2009sk} (BGP$^2$S), measurements  of the prompt cross sections for $X$ and $\psi(2S)$
by the CDF Collaboration were used to obtain an estimate of the product $\sigma\,\mathrm{Br}$ 
of the prompt cross section for $X$ and its branching fraction into $J/\psi\, \pi^+\pi^-$ in the region
with rapidity $|y| < 0.6$ and transverse momentum $p_T > 5$~GeV:
\begin{equation}
\mathrm{Tevatron:}~~\sigma[X(3872)]\, \mathrm{Br} [X \to J/\psi\, \pi^+\pi^-] \approx (3.1 \pm 0.7)~ \mathrm{nb}.
\label{sigmaBr:CDF}
\end{equation}
A loose lower bound on the cross section for $X$ can be obtained by using $\mathrm{Br} < 1$.

The inclusive production of the $X$ in $pp$ collisions at the LHC has been studied by the LHCb \cite{Aaij:2011sn},
CMS \cite{Chatrchyan:2013cld}, and ATLAS \cite{Aaboud:2016vzw} collaborations.
The CMS collaboration measured the product $\sigma\,\mathrm{Br}$ of the prompt cross section for $X$
and its branching fraction into $J/\psi\, \pi^+\pi^-$ for $X$ with $|y|< 1.2$ and
 $10\, \mathrm{GeV} < p_T < 30\, \mathrm{GeV}$  at center-of-mass energy 7~TeV \cite{Chatrchyan:2013cld}:
\begin{equation}
\mathrm{LHC:}~~\sigma[X(3872)]\, \mathrm{Br} [X \to J/\psi\, \pi^+\pi^-] = (1.06 \pm 0.11\pm 0.15 )~ \mathrm{nb}.
\label{sigmaBr:CMS}
\end{equation}
The prompt fraction was measured to be 
about 74\% in the range   $10\, \mathrm{GeV} < p_T < 50\, \mathrm{GeV}$ \cite{Chatrchyan:2013cld}.

In Ref.~\cite{Braaten:2019ags}, we derived upper and lower bounds on the branching fraction Br 
for the $X$ bound state to decay into $J/\psi\, \pi^+\pi^-$:
\begin{equation}
4\%  <  \mathrm{Br} [X \to J/\psi\, \pi^+\pi^-] < 33\%.
\label{BRXtopsipipi}
\end{equation}
The loose lower bound $\mathrm{Br} > 4\%$ is derived from a recent measurement by the BaBar collaboration
of the inclusive branching fraction for $B^+$ into $K^+$ plus the  $X$ resonance feature \cite{Wormser}. 
The upper bound $\mathrm{Br} < 33\%$ is derived from measurements of branching ratios 
of $J/\psi\, \pi^+\pi^-$ over other short-distance decay modes of the $X$.

Given the results for $\sigma\,\mathrm{Br}$ in Eqs.~\eqref{sigmaBr:CDF} and \eqref{sigmaBr:CMS},
a constraint on the branching fraction into $J/\psi\, \pi^+\pi^-$ gives constraints on the prompt cross sections.
A lower bound on the prompt cross section $\sigma$ can be estimated by decreasing the central value of
$\sigma\,\mathrm{Br}$ by 1 standard deviation  and then dividing it by the upper bound on Br in  Eq.~\eqref{BRXtopsipipi}.
A loose upper bound on $\sigma$ can be estimated by increasing the central value of
$\sigma\,\mathrm{Br}$  by 1 standard deviation  and then dividing it by the lower bound on Br in  Eq.~\eqref{BRXtopsipipi}.
Using the estimate in Eq.~\eqref{sigmaBr:CDF}, the bounds on the prompt cross section at the Tevatron 
with  $|y| < 0.6$ and $p_T > 5$~GeV are
\begin{equation}
\mathrm{Tevatron:}~~7.3~\mathrm{nb} < \sigma[X(3872)] < 95~\mathrm{nb}.
\label{sigma:Tevatron}
\end{equation}
Using the measurement in Eq.~\eqref{sigmaBr:CMS}, the bounds on the prompt cross section at the LHC 
with  $|y|< 1.2$ and  $10\, \mathrm{GeV} < p_T < 30\, \mathrm{GeV}$ are
\begin{equation}
\mathrm{LHC:}~~2.6~\mathrm{nb} < \sigma[X(3872)] < 31~\mathrm{nb}.
\label{sigma:LHC}
\end{equation}
In both Eqs.~\eqref{sigma:Tevatron} and \eqref{sigma:LHC},
the loose upper bound is  more than 10 times larger than the lower bound.

In Ref.~\cite{Albaladejo:2017blx}, ranges of prompt cross sections $\sigma$ 
for $X$ at the Tevatron and at the LHC were obtained by considering branching fractions in the 
range  $2.7\%  <  \mathrm{Br}  < 8.3\%$ \cite{Guo:2014sca}.
The lower ends of their ranges were about  5 times larger than our lower bounds 
on the cross sections in Eqs.~\eqref{sigma:Tevatron} and \eqref{sigma:LHC}.  
The differences come  primarily from two sources.
First, the lower end of their range for  $\sigma$ was obtained from the central value
 of $\sigma\,\mathrm{Br}$ rather than from the value lower by 1 standard deviation.
Second, the upper end of their range for Br was about 4 times smaller 
than our upper bound in  Eq.~\eqref{BRXtopsipipi}.
The smaller upper bound on Br was obtained by assuming that 
measurements of the branching fraction of $X$ into $D^0 \bar D^{*0}$ are dominated by
the decay of the $X$ resonance into $D^0 \bar D^0 \pi^0$ and $D^0 \bar D^0 \gamma$
below the $D^0 \bar D^{*0}$ threshold and have a negligible contribution 
from the threshold enhancement in the production of $D^0 \bar D^{*0}$ above the threshold.
This assumption is contradicted by measurements of the width of the $X$ from the $D^0 \bar D^{*0}$ decay  mode,
which are significantly larger than the upper bound on the width obtained by the Belle collaboration from the 
$J/\psi\, \pi^+\pi^-$ decay mode \cite{Choi:2011fc}.


\section{ Upper bound on cross section for $\bm{X}$}
\label{sec:Upperbound}

In this section, we present the  upper bound on the inclusive prompt cross section for
producing $X(3872)$  in Ref.~\cite{Bignamini:2009sk}.
We explain why the derivation of  the upper bound requires the charm-meson pair cross section 
to be integrated up to a relative momentum of order $m_\pi$ instead  of order $\gamma_X$,
as apparently assumed in  Ref.~\cite{Bignamini:2009sk}.

If $X$ is a charm-meson molecule with the flavor structure in Eq.~\eqref{Xflavor},
the inclusive cross section for producing $X$ can  be expressed in terms of the same amplitudes
as those in the inclusive cross sections for producing $D^{*0} \bar D^0$ and $D^0 \bar D^{*0}$ \cite{Bignamini:2009sk}.
The inclusive cross sections for  producing $D^{*0} \bar D^0$ and $D^0 \bar D^{*0}$
with small relative momentum $\bm{k}$ in the charm-meson-pair rest frame
and the inclusive cross section for  producing $X$ can be expressed as
\begin{subequations}
\begin{eqnarray}
d\sigma[D^{*0} \bar D^0] &=&
\frac{1}{\mathrm{flux}} \sum_y \int d\Phi_{(D^*\bar D)+y}
\Big|  \mathcal{A}_{D^{*0} \bar D^0+y}(\bm{k}) \Big|^2 
\frac{d^3k}{(2 \pi)^32 \mu},
\label{sigmaDstarD}
\\
d\sigma[D^0 \bar D^{*0}] &=&
\frac{1}{\mathrm{flux}} \sum_y \int d\Phi_{(D^*\bar D)+y}
\Big|  \mathcal{A}_{D^{0} \bar D^{*0}+y}(\bm{k}) \Big|^2 
 \frac{d^3k}{(2 \pi)^32 \mu},
\label{sigmaDDstar}
\\
d\sigma[X(3872)] &=&
\frac{1}{\mathrm{flux}} \sum_y \int d\Phi_{(D^*\bar D)+y}
\left|  \int \!\!\frac{d^3k}{(2 \pi)^3}  \psi_X(k) 
\frac{ \mathcal{A}_{D^{*0} \bar D^0+y}(\bm{k}) +  \mathcal{A}_{D^{0} \bar D^{*0}+y}(\bm{k})}{\sqrt2}\right|^2 
\frac{1}{2\mu},
\nonumber\\
\label{sigmaX}
\end{eqnarray}
\label{sigmaDstarD,DDstar,X}%
\end{subequations}
where $\mu$ is the reduced mass of $D^{*0} \bar D^0$.
The sums over $y$ are over all the additional particles that can be produced.
The  amplitudes that appear in the cross section for $X$ in Eq.~\eqref{sigmaX} are the charge-conjugation-even 
superpositions of the amplitudes for producing $D^{*0} \bar D^0+y$ and $D^{0} \bar D^{*0}+y$.
The momentum-space wavefunction for the $X$ in Eq.~\eqref{sigmaX} is normalized so 
$ \int (d^3k/(2 \pi)^3)\, |\psi_X(k)|^2 = 1$.
The differential phase space $d\Phi_{(D^*\bar D)+y}$  is that for a composite particle 
 denoted by $(D^*\bar D)$ with mass $M_{*0} + M_0$ plus the additional particles $y$.  
The mass of $X$ is sightly smaller than $M_{*0} + M_0$ and the invariant mass of a charm-meson pair
is larger than $M_{*0} + M_0$, but the differences  in the phase space integrals  are negligible.
Factors of 3 from the sums over the spin states of $D^{*0}$ or $\bar D^{*0}$ or $X$
are absorbed into the amplitudes $ \mathcal{A}$.
The phase-space integrals in Eqs.~\eqref{sigmaDstarD,DDstar,X} are over the 3-momenta of the additional particles $y$,
 but the cross sections remain differential in the 3-momentum $\bm{P}$ of $(D^*\bar D)$.
 Thus the $D^{*0} \bar D^0$ and $D^0 \bar D^{*0}$ cross sections in Eqs.~\eqref{sigmaDstarD} and \eqref{sigmaDDstar}
 are differential in both $\bm{P}$ and $\bm{k}$, while the $X$ cross section in Eq.~\eqref{sigmaX}
is differential only in  $\bm{P}$.

In the expression for the $X$ cross section in Eq.~\eqref{sigmaX}, there are  interference terms 
between the amplitudes for producing $D^{*0} \bar D^0+y$ and $D^0 \bar D^{*0}+y$.
The interference terms are positive for some sets of additional final-state particles $y$ and negative for others.
In high-energy hadron collisions, there are dozens or even hundreds of additional particles.
The sum over the many additional particles $y$ gives cancellations that suppress the interference terms. 
The $X$ cross section in Eq.~\eqref{sigmaX} then reduces to the sum of a $D^{*0} \bar D^0$ term 
and a $D^0 \bar D^{*0}$ term.  At large transverse momentum, the hadronization of a $c \bar c$ pair 
is equally likely to produce $D^{*0} \bar D^0$ and $D^0 \bar D^{*0}$,
because the probability of a light quark or antiquark from  a colliding hadron
to become a constituent of one of the charm mesons is very small.
The $D^{*0} \bar D^0$ term and the $D^0 \bar D^{*0}$ term should therefore be equal, 
and the expression for the $X$ cross section can be reduced to 
\begin{equation}
d\sigma[X(3872)] =
\frac{1}{\mathrm{flux}} \sum_y \int d\Phi_{(D^*\bar D)+y}
\left|  \int \!\!\!\frac{d^3k}{(2 \pi)^3}  \psi_X(k) 
 \mathcal{A}_{D^{*0} \bar D^0+y}(\bm{k}) \right|^2 
\frac{1}{2\mu}.
\label{sigmaX-1}
\end{equation}

The authors of  Ref.~\cite{Bignamini:2009sk} (BGP$^2$S)
derived a theoretical upper bound on  the cross section for producing $X$
in terms of a  cross section for producing the charm meson pair $D^{*0} \bar D^0$.
To derive their upper bound, BGP$^2$S first restricted the integral over the relative momentum 
in Eq.~\eqref{sigmaX-1} to a region $|\bm{k}|<k_\mathrm{max}$ in which $\psi_X(k)$ differs significantly from 0.
They then applied the Schwarz inequality to that integral:
\begin{equation}
d\sigma[X(3872)] \le
\int\bm{'}\!\!\!\frac{d^3k}{(2\pi)^3} \, \big|\psi_X(k)\big|^2 \cdot
\frac{1}{\mathrm{flux}} \sum_y \int d\Phi_{(D^*\bar D)+y}
\int\bm{'}\!\!\!\frac{d^3k}{(2\pi)^3} \big|\mathcal{A}_{D^{*0} \bar D^0+y}(\bm{k})\big|^2 
 \frac{1}{2\mu},
\label{sigmaXle}
\end{equation}
where the primes on the integrals indicate restrictions to $|\bm{k}|<k_\mathrm{max}$.
The first factor on the right side of Eq.~\eqref{sigmaXle} is the probability for the constituents of the $X$ 
to have relative momentum less than $k_\mathrm{max}$. Since this probability is less than 1,
 they obtained the inequality
\begin{equation}
\sigma[X(3872)]  < \sigma\big[D^{*0} \bar D^0(k< k_\mathrm{max})\big] .
\label{sigmaX<sigmaDDbar}
\end{equation}
The validity of this inequality hinges on the validity of restricting the integral 
in Eq.~\eqref{sigmaX-1}  to the region $|\bm{k}|<k_\mathrm{max}$.

In Ref.~\cite{Bignamini:2009sk}, BGP$^2$S did not give an unambiguous prescription 
for the maximum momentum $k_\mathrm{max}$ in the  inequality in Eq.~\eqref{sigmaX<sigmaDDbar}.
They quoted the difference $E_X$ between the mass of the $X$ and the energy 
of the $D^{*0} \bar D^0$ scattering threshold at that time as $E_X = -0.25 \pm 0.40$~MeV.
The central value corresponds to binding momentum $\gamma_X=22$~MeV,
and the value lower by $1\sigma$ corresponds to  $\gamma_X=35$~MeV.
In Ref.~\cite{Bignamini:2009sk}, BGP$^2$S chose $k_\mathrm{max}$ 
in the inequality in Eq.~\eqref{sigmaX<sigmaDDbar} to be 35~MeV.
In a subsequent paper Ref.~\cite{Esposito:2017qef},
whose authors included most of those of Ref.~\cite{Bignamini:2009sk},
an updated estimate $k_\mathrm{max}=20$~MeV was given.
This is close to the  binding momentum $\gamma_X=18$~MeV from the 
value  of $E_X$ that is $1\sigma$ below the central value  in Eq.~\eqref{eq:deltaMX}.
The choice for $k_\mathrm{max}$ in both papers is consistent with the assumption that $k_\mathrm{max}$
is approximately $\gamma_X$, although this assumption was not stated explicitly in Ref.~\cite{Bignamini:2009sk}.

The conclusions of Ref.~\cite{Bignamini:2009sk} were challenged in Ref.~\cite{Artoisenet:2009wk},
which argued that the appropriate choice of $k_\mathrm{max}$ 
in the upper bound in Eq.~\eqref{sigmaX<sigmaDDbar} is of order $m_\pi$ instead of order $\gamma_X$.
If short-range interactions produce an S-wave bound state close to a scattering threshold,
the momentum-space wavefunction in the momentum region below the inverse range
has the universal form in Eq.~\eqref{psiX-k}.
The normalization integral of the probability density $|\psi_X(k)|^2$ is dominated by $k$ of order $\gamma_X$.
However the integral over $\bm{k}$  in Eq.~\eqref{sigmaX},
whose integrand  has only one factor of $\psi_X(k)$, is not dominated by $k$ of order $\gamma_X$.
It  has significant contributions from the region extending up to $k$ of order $m_\pi$,
which is where the wavefunction $\psi_X(k)$ begins to fall faster than $1/k^2$.
Thus the derivation of the upper bound in Ref.~\cite{Bignamini:2009sk} requires  
$k_\mathrm{max}$ to be of order $m_\pi$ instead of order $\gamma_X$, 
as was apparently assumed in Ref.~\cite{Bignamini:2009sk}.

In Ref.~\cite{Bignamini:2009sk}, BGP$^2$S estimated the  cross section for charm-meson pairs 
with relative momentum $k$ at  the Tevatron using the event generators Herwig and Pythia 
to produce hadronic final states from $2\to 2$ parton processes, primarily $g g \to gg$.
This extremely inefficient method gave distributions at small $k$ with the behavior $k \, dk$.
Their estimate for the theoretical upper bound on the prompt cross section for $X$ at the Tevatron 
obtained by inserting
$k_\mathrm{max}=35$~MeV into Eq.~\eqref{sigmaX<sigmaDDbar} was 0.07~nb using Herwig and 0.11~nb using Pythia. 
These cross sections are about 30 times smaller than the loose lower bound
of 3.1~nb given by the right side of Eq.~\eqref{sigmaBr:CDF}.
BGP$^2$S concluded that if the $X$ was a weakly bound charm-meson molecule,
its formation from charm mesons at the rate  observed at the Tevatron would be unlikely.
Given that their $D^{*0} \bar D^0$ cross section scaled as $k_\mathrm{max}^2$, 
the value of $k_\mathrm{max}$ would have to be larger than about 280~MeV for the $D^{*0} \bar D^0$ cross section calculated using Pythia to be above the 
lower bound on the $X$ cross section at the Tevatron in  Eq.~\eqref{sigma:Tevatron}.

In Ref.~\cite{Artoisenet:2009wk}, the  cross section for charm-meson pairs with small relative momentum $k$ 
at  the Tevatron was estimated using the event generator Pythia  to produce hadronic final states from
the $2\to 3$ parton process $gg \to c \bar c g$. 
The distribution had the behavior $k^2\, dk$ in the region $k < m_\pi$.
This behavior should be more accurate than the behavior $k\, dk$ obtained in Ref.~\cite{Bignamini:2009sk},
because the event generator, which has not been tuned to reproduce distributions in $k$,
 plays a smaller role in generating the distributions.
 Their estimate for the charm-meson-pair cross section at the Tevatron 
integrated up to a relative momentum $k_\mathrm{max}$ was
\begin{equation}
\mathrm{Tevatron:}~~\sigma_\mathrm{naive}\big[D^{*0} \bar D^0(k< k_\mathrm{max})\big] 
\approx 0.03~\mathrm{nb} \left( \frac{k_\mathrm{max}}{35~\mathrm{MeV}} \right)^3.
\label{sigmaDD:Tevatron}
\end{equation}
The subscript ``naive'' emphasizes that the cross section calculated using an event generator
does not take into account the effects of the $X$ resonance.
The theoretical upper bound in Eq.~\eqref{sigmaX<sigmaDDbar} is greater than the loose lower bound of 3.1~nb given 
by  the right side of Eq.~\eqref{sigmaBr:CDF} if $k_\mathrm{max}$ is greater than about 160~MeV.
The authors of Ref.~\cite{Artoisenet:2009wk} concluded that the upper bound in Eq.~\eqref{sigmaX<sigmaDDbar} 
with $k_\mathrm{max}$ of order $m_\pi$ was compatible with the observed prompt cross section for $X$ at the Tevatron. 
The upper and lower bounds on the $X$ cross section at the Tevatron in  Eq.~\eqref{sigma:Tevatron}
take into account the  bounds on the branching fraction for 
$X \to J/\psi\, \pi^+\pi^-$ in Eq.~\eqref{BRXtopsipipi}.
The value of $k_\mathrm{max}$ would have to be larger than 220~MeV for 
 the naive $D^{*0} \bar D^0$ cross section to be above the
lower bound on the $X$ cross section at the Tevatron in  Eq.~\eqref{sigma:Tevatron}.
This value of $k_\mathrm{max}$ is also compatible with the scale $m_\pi$.

The conclusions of Ref.~\cite{Bignamini:2009sk} were  also challenged in Ref.~\cite{Albaladejo:2017blx}.
They used an effective field theory with  ultraviolet cutoff $\Lambda$ 
in which $X$ is treated as a charm-meson molecule.
They calculated the  inclusive prompt cross sections for producing $X$ 
in $p \bar p$ collisions at the Tevatron and in $p p$ collisions at the LHC
using the event generators Herwig and Pythia to calculate the production rate of $D^* \bar D$ at short distances 
and using the effective field theory to calculate the formation rate of the $X$ at long distances.
With $\Lambda = 100$~MeV, 
their cross sections using Pythia were 0.05~nb at the Tevatron  and 0.04~nb at the LHC,
which are much smaller than the results from CDF and CMS in  Eqs.~\eqref{sigmaBr:CDF} and \eqref{sigmaBr:CMS}.
Their cross sections were compatible with the results 
from CDF and CMS for $\Lambda$  in the range from 500~MeV to 1000~MeV.
For these large ultraviolet cutoffs, the contributions from the charged-charm-meson-pair channels 
$D^{*+} D^-$ and $D^{+} D^{*-}$ were larger than those from the 
$D^{*0} \bar D^0$ and $D^{0} \bar D^{*0}$ channels by about a factor of 2.

The analysis in Ref.~\cite{Albaladejo:2017blx} was rejected in Ref.~\cite{Esposito:2017qef},
whose authors included most of those of Ref.~\cite{Bignamini:2009sk}.
They argued that $k_\mathrm{max}$ must be determined 
``independently of any educated guesses on the explicit form'' of $\psi_X(k)$.
They did not address the issue that their derivation of the upper bound with $k_\mathrm{max}$
approximately equal to 
$\gamma_X$ fails for the explicit wavefunction in Eq.~\eqref{psiX-k}.


\section{Estimate of  cross section for $\bm{X}$}
\label{sec:Equality}

In this section, we derive an equality between the $X(3872)$ cross section  
and a charm-meson-pair cross section that takes into account the threshold enhancement 
produced by the $X$ resonance.
We also present an order-of-magnitude estimate of the $X$ cross section
in terms of a naive charm-meson-pair cross section that does not take into account the threshold enhancement.

Expressions for the cross sections for producing charm meson pairs and for producing $X$
 in Eqs.~\eqref{sigmaDstarD,DDstar,X} that take into account 
the $X$ resonance were presented in Ref.~\cite{Artoisenet:2009wk}.
The cross sections were expressed in factored forms,
with long-distance factors that involve the binding momentum $\gamma_X$ and with 
short-distance factors that involve only momentum scales of order $m_\pi$ or larger.
The amplitude for producing $D^{*0} \bar D^0+y$  in Eq.~\eqref{sigmaDstarD} 
can be decomposed into charge-conjugation even ($C=+$) and charge-conjugation odd ($C=-$) components.
The $C=+$ component  is enhanced by the $X$ resonance.
If the nonresonant $C=-$ component is neglected, the amplitude for producing  $D^{*0} \bar D^0+y$
can be expressed as a product of the $C=+$ component of a short-distance
amplitude and a resonance factor that depends on $\gamma_X$:
\begin{equation}
 \mathcal{A}_{D^{*0} \bar D^0+y}(\bm{k})  = \frac{1}{\sqrt2} \left(
\frac{\mathcal{A}^{\mathrm{s.d.}}_{D^{*0} \bar D^0+y} + \mathcal{A}^{\mathrm{s.d.}}_{D^0 \bar D^{*0}+y}}{\sqrt2} \right)
\, \frac{\Lambda}{-\gamma_X-ik}.
\label{amp-DstarDbar}
\end{equation}
The expression for the corresponding amplitude $\mathcal{A}_{D^0 \bar D^{*0}+y}(\bm{k})$  is identical.
The short-distance amplitudes $\mathcal{A}^{\mathrm{s.d.}}_{D^{*0} \bar D^0+y}$
and $\mathcal{A}^{\mathrm{s.d.}}_{D^0 \bar D^{*0}+y}$
are independent of the momentum if $\bm{k}$ is small compared to  $m_\pi$.
The constant $\Lambda$ in the numerator of the resonance factor should be of order $m_\pi$.
The only dependence on the small momentum $\gamma_X$ is in the denominator of the resonance factor.
Since $\Lambda \gg \gamma_X$, the absolute value of the resonance factor is approximately 1 at $k = \Lambda$,
so $\Lambda$  can be interpreted as the momentum scale 
where  the amplitude becomes comparable in magnitude to the amplitude in the absence of the resonance.
The resonance factor in  Eq.~\eqref{amp-DstarDbar} produces a threshold enhancement in the cross section.
The differential cross section $d\sigma/dE$ in the kinetic energy $E$ of $D^{*0} \bar D^0$ in
the $D^{*0} \bar D^0$ center-of-momentum (CM) frame should have a local minimum above the threshold enhancement.
A simple physical interpretation of $\Lambda$  is that the kinetic energy $E$ at the  local minimum is
roughly $\Lambda^2/2\mu$.

The factorization formula for the $D^{*0} \bar D^0$ cross section can be obtained simply 
by inserting the amplitude in Eq.~\eqref{amp-DstarDbar} into Eq.~\eqref{sigmaDstarD}.
The factorization formula for the $X$ cross section  cannot be obtained so simply.
If the universal wavefunction in Eq.~\eqref{psiX-k} is inserted into Eq.~\eqref{sigmaX},
 the momentum integral is logarithmically ultraviolet divergent.
The factorization formula for the $X$ cross section can be obtained instead by requiring the sum of the cross sections 
for producing $X$ and the cross sections for producing $D^{*0} \bar D^0$ and $D^0 \bar D^{*0}$
integrated over $\bm{k}$ to be consistent with the optical theorem in Eq.~\eqref{Imf-E}.
The resulting factorization formulas for the inclusive cross sections are
\begin{subequations}
\begin{eqnarray}
d\sigma[D^{*0} \bar D^0] &=&
\frac{1}{\mathrm{flux}} \sum_y \int d\Phi_{(D^*\bar D)+y}
\left| \mathcal{A}^{\mathrm{s.d.}}_{D^{*0} \bar D^0+y} + \mathcal{A}^{\mathrm{s.d.}}_{D^0 \bar D^{*0}+y} \right|^2 
\frac{\Lambda^2}{\gamma_X^2+k^2} \frac{d^3k}{(2 \pi)^3 8 \mu},
\label{factor-DstarDbar}
\\
d\sigma[D^0 \bar D^{*0}] &=&
\frac{1}{\mathrm{flux}} \sum_y \int d\Phi_{(D^*\bar D)+y}
\left| \mathcal{A}^{\mathrm{s.d.}}_{D^{*0} \bar D^0+y} + \mathcal{A}^{\mathrm{s.d.}}_{D^0 \bar D^{*0}+y} \right|^2 
\frac{\Lambda^2}{\gamma_X^2+k^2}  \frac{d^3k}{(2 \pi)^3 8 \mu},
\label{factor-DDbarstar}
\\
d\sigma[X(3872)] &=&
\frac{1}{\mathrm{flux}} \sum_y \int d\Phi_{(D^*\bar D)+y}
\left| \mathcal{A}^{\mathrm{s.d.}}_{D^{*0} \bar D^0+y} + \mathcal{A}^{\mathrm{s.d.}}_{D^0 \bar D^{*0}+y} \right|^2 
\frac{\Lambda^2\gamma_X}{8\pi\mu}.
\label{factor-X}
\end{eqnarray}
\label{factor-DstarD,X}%
\end{subequations}
The differential cross sections for $D^{*0} \bar D^0$ and $D^{0} \bar D^{*0}$ in Eqs.~\eqref{factor-DstarDbar} 
and \eqref{factor-DDbarstar} should be good approximations up to relative momentum $k$ of about $\Lambda$.

The short distance factors in Eqs.~\eqref{factor-DstarD,X} 
can be eliminated to obtain an expression for the $D^{*0} \bar D^0$ cross section 
 in terms of the $X$ cross section:
\begin{equation}
d\sigma[D^{*0} \bar D^0]  = 
d\sigma[X(3872)] \,\frac{\pi/\gamma_X}{\gamma_X^2+k^2}  \frac{d^3k}{(2 \pi)^3}.
\label{dsigmaDstarD-X}
\end{equation}
This relation is analogous to relations between cross sections for the production of the deuteron bound state
 with large momentum transfer and cross sections for the production of proton-neutron pairs \cite{Faeldt:1996na}.
The integral of this $D^{*0} \bar D^0$ cross section over the region $|\bm{k}| < k_\mathrm{max}$ is
\begin{eqnarray}
\sigma[D^{*0} \bar D^0(k<k_\mathrm{max})]  &=& \sigma[X(3872)] 
 \frac{k_\mathrm{max}/\gamma_X -\arctan(k_\mathrm{max}/\gamma_X)}{2\pi}.
\label{sigmaDstarD-X}
\end{eqnarray}
There is a value of $k_\mathrm{max}$ such that the integrated cross section is equal to that for $X$:
$k_\mathrm{max} = 7.73\,\gamma_X$.
The resulting equality between the $X$ cross section and a $D^{*0} \bar D^0$ cross section is
\begin{equation}
\sigma[X(3872)] = \sigma[D^{*0} \bar D^0(k<7.73\,  \gamma_X)] .
\label{sigmaX=sigmaDstarD}
\end{equation}
The right side of this equality can also be expressed as the sum of the $D^{*0} \bar D^0$ 
and $D^0 \bar D^{*0}$ cross sections
integrated up to a smaller maximum value of $k$:
\begin{equation}
\sigma[X(3872)] = 2\, \sigma[D^{*0} \bar D^0(k<4.49\,  \gamma_X)] .
\label{sigmaX=sigmaDstarDx2}
\end{equation}
The equalities in Eqs.~\eqref{sigmaX=sigmaDstarD} and \eqref{sigmaX=sigmaDstarDx2}
are equivalent provided $\gamma_X$ is sufficiently small.
As a reasonable condition for the validity of the equality, we require $k_\mathrm{max}$  to be
smaller than the inverse range $m_\pi$.
If this condition is applied to the equalities in Eqs.~\eqref{sigmaX=sigmaDstarD} and \eqref{sigmaX=sigmaDstarDx2},
it requires the binding energy $|E_X|$ to be less than about 
0.2~ MeV and 0.5~MeV, respectively.

The equality in Eq.~\eqref{sigmaX=sigmaDstarD} is consistent with the upper bound in 
Eq.~\eqref{sigmaX<sigmaDDbar} for any value of $k_\mathrm{max}$ greater than $7.73\,\gamma_X$.
However the derivation of the upper bound in  Eq.~\eqref{sigmaX<sigmaDDbar}  requires 
$k_\mathrm{max}$ to be order $m_\pi$, because the derivation must allow for the
 possibility that the wavefunction for $X$ has the form in Eq.~\eqref{psiX-k}.
The equality in Eq.~\eqref{sigmaX=sigmaDstarD} is incompatible with the upper bound in 
Eq.~\eqref{sigmaX<sigmaDDbar} if $k_\mathrm{max}$ is taken to be approximately $\gamma_X$,
as was apparently  assumed in Ref.~\cite{Bignamini:2009sk}.

The equality between the cross sections for $X$ and $D^{*0} \bar D^0$ in Eq.~\eqref{sigmaX=sigmaDstarD}
assumes  the $D^{*0} \bar D^0$ cross section has the threshold enhancement  
from the $X$ resonance. 
If an event generator such as Herwig or Pythia is used to estimate
the charm-meson-pair cross section, the equality in Eq.~\eqref{sigmaX=sigmaDstarD} cannot be used
because the event generator is not informed about the resonance. 
The naive cross section for producing $D^{*0} \bar D^0$ can be obtained from Eq.~\eqref{sigmaDstarD}
by replacing the amplitude $\mathcal{A}_{D^{*0} \bar D^0+y}(\bm{k})$
by the short-distance amplitude $ \mathcal{A}^{\mathrm{s.d.}}_{D^{*0} \bar D^0+y}$:
\begin{equation}
d\sigma[D^{*0} \bar D^0]_\mathrm{naive} \approx
\frac{1}{\mathrm{flux}} \sum_y \int d\Phi_{(D^*\bar D)+y}
\left| \mathcal{A}^{\mathrm{s.d.}}_{D^{*0} \bar D^0+y}\right|^2  \frac{d^3k}{(2 \pi)^3 2 \mu}.
\label{factor-DstarDbar-naive}
\end{equation}
This naive $D^{*0} \bar D^0$ cross section integrated over the region $|\bm{k}| < k_\mathrm{max}$ 
scales like $k_\mathrm{max}^3$, in agreement with  the charm meson cross section 
calculated using an event generator  in Ref.~\cite{Artoisenet:2009wk}.
In the expression for the $X$ cross section in Eq.~\eqref{factor-X},
the sum over the many additional particles $y$  give cancellations that suppress 
the interference terms between the amplitudes 
$\mathcal{A}^{\mathrm{s.d.}}_{D^{*0} \bar D^0+y}$ and $\mathcal{A}^{\mathrm{s.d.}}_{D^0 \bar D^{*0}+y}$. 
In a high energy hadron collider, the
production rates for $D^{*0} \bar D^0$ and $D^0 \bar D^{*0}$  at large transverse momentum
should be equal, because the light quarks in the charm mesons are unlikely to come from the colliding hadrons.
The $D^{*0} \bar D^0$ and $D^0 \bar D^{*0}$ contributions should therefore be equal, so
the cross section reduces to
\begin{equation}
d\sigma[X(3872)] =
\frac{1}{\mathrm{flux}} \sum_y \int d\Phi_{(D^*\bar D)+y}
\left| \mathcal{A}^{\mathrm{s.d.}}_{D^{*0} \bar D^0+y}\right|^2 
\frac{\Lambda^2\gamma_X}{4\pi\mu}.
\label{factor-X-interference}
\end{equation}
The short distance factor in the expression for the naive $D^{*0} \bar D^0$ cross section in 
Eq.~\eqref{factor-DstarDbar-naive} can then be eliminated  in favor
 of the $X$ cross section using Eq.~\eqref{factor-X-interference}:
\begin{equation}
d\sigma[D^{*0} \bar D^0]_\mathrm{naive}  \approx
d\sigma[X(3872)] \,\frac{2 \pi/\gamma_X}{\Lambda^2}   \frac{d^3k}{(2\pi)^3}.
\label{dsigmaDstarD-X:naive}
\end{equation}
Note that this is larger by a factor of 2 than the cross section obtained from the 
equality in Eq.~\eqref{dsigmaDstarD-X} by replacing $\gamma_X^2+k^2$ in the denominator by $\Lambda^2$.
If the naive $D^{*0} \bar D^0$ cross section in Eq.~\eqref{dsigmaDstarD-X:naive} is integrated 
over the region $|\bm{k}| < k_\mathrm{max}$,
there is a value of $k_\mathrm{max} $ for which the integrated cross section is equal  to the $X$ cross section:
\begin{equation}
\sigma[X(3872)] \approx  \sigma_\mathrm{naive}[D^{*0} \bar D^0(k<(3 \pi \Lambda^2 \gamma_X)^{1/3}] .
\label{sigmaX=sigmaDstarDnaive}
\end{equation}
This result can be used to obtain an order of magnitude estimate of the $X$ cross section
using naive charm-meson-pair cross sections calculated using a Monte Carlo event generator.
Since $\Lambda$ is order $m_\pi$, this estimate is compatible with the upper bound 
in Eq.~\eqref{sigmaX<sigmaDDbar} with $k_\mathrm{max}$ of order $m_\pi$.

In Ref.~\cite{Artoisenet:2009wk}, the estimate of the $X$ cross section
in terms of a naive charm-meson-pair cross section was expressed in the form
\begin{equation}
\sigma[X(3872)] \approx  
\sigma_\mathrm{naive}[D^{*0} \bar D^0(k<\Lambda) ] \frac{6 \pi \gamma_X}{\Lambda}.
\label{sigmaX=sigmaDstarDnaivePA}
\end{equation}
The naive $D^{*0} \bar D^0$ cross section integrated up to a relative momentum $k_\mathrm{max}$
that was calculated using the Pythia event generator in Ref.~\cite{Artoisenet:2009wk}
scales as $k_\mathrm{max}^3$.
The estimate in Eq.~\eqref{sigmaX=sigmaDstarDnaivePA} therefore differs from that in 
Eq.~\eqref{sigmaX=sigmaDstarDnaive} only by a multiplicative factor of 2.
Thus the estimate  in Eq.~\eqref{sigmaX=sigmaDstarDnaive} is essentially  just a convenient repackaging of 
 the estimate from Ref.~\cite{Artoisenet:2009wk} in Eq.~\eqref{sigmaX=sigmaDstarDnaivePA}.

The estimate of the $X$ cross section in Eq.~\eqref{sigmaX=sigmaDstarDnaive}
depends on the combination $(\Lambda^2 \gamma_X)^{1/3}$ of unknown parameters.
If we use the lower bound on the $X$ cross section at the Tevatron in Eq.~\eqref{sigma:Tevatron}
and the naive estimate of the $D^{*0} \bar D^0$ cross section  at the Tevatron in Eq.~\eqref{sigmaDD:Tevatron},
we get the lower bound  $(\Lambda^2 \gamma_X)^{1/3} > 100$~MeV.
The order-of-magnitude estimate of this combination of parameters obtained by inserting $\Lambda= m_\pi$
and $\gamma_X = 18$~MeV is $(\Lambda^2 \gamma_X)^{1/3} \sim 70$~MeV.
We conclude that the estimate of the $X$ cross section at the Tevatron in Eq.~\eqref{sigmaX=sigmaDstarDnaivePA}
is compatible with the observed cross section for some value of $\Lambda$ of order $m_\pi$
if the binding energy of the $X$ is roughly 0.17 MeV.

There does not seem to be any calculation using event generators in the literature of the 
naive cross section for producing $D^{*0} \bar D^0$ with small relative momentum at the LHC.
If there was such a calculation for a single small value of $k_\mathrm{max}$,
the cross section as a function of $k_\mathrm{max}$ could be obtained simply by 
assuming it scales as $k_\mathrm{max}^3$.
Eq.~\eqref{sigmaX=sigmaDstarDnaive} would then  give an  estimate of the cross section for producing $X$ 
at the LHC that could be compared with the measured value in Eq.~\eqref{sigma:LHC}.
Given the large uncertainty in $k_\mathrm{max} = (3 \pi \Lambda^2 \gamma_X)^{1/3}$
and the fact that the estimate of the cross section scales as $k_\mathrm{max}^3$,
there is little doubt that the estimate would be compatible with the observed  cross section.


\section{Summary and Discussion}
\label{sec:Summary}

We have discussed the inclusive prompt production of the $X(3872)$ at high energy hadron colliders under the assumption 
that the $X$ is a weakly bound charm-meson molecule with the particle content in Eq.~\eqref{Xflavor}.
We considered the production of $X$ through the creation of
its constituents $D^{*0} \bar D^0$ or $D^0 \bar D^{*0}$ at short distances
of order $1/m_\pi$ or smaller. The formation of the $X$ proceeds on longer distance scales
of order $1/\gamma_X$, where $\gamma_X$ is the binding momentum.  

The theoretical upper bound on the cross section for producing $X$ in Eq.~\eqref{sigmaX<sigmaDDbar}
was derived in Ref.~\cite{Bignamini:2009sk}.
It is given by the $D^{*0} \bar D^0$ cross section integrated up to relative momentum $k_\mathrm{max}$.
The authors did not give any clear prescription for $k_\mathrm{max}$,
but their numerical value for $k_\mathrm{max}$  was consistent with  it being
approximately equal to  $\gamma_X$.
In Ref.~\cite{Artoisenet:2009wk}, it was pointed out that the derivation of the upper bound in
Eq.~\eqref{sigmaX<sigmaDDbar} actually requires $k_\mathrm{max}$ to be of order $m_\pi$.
A specific example of a wavefunction for which the derivation of the upper bound requires
$k_\mathrm{max} \gg \gamma_X$ is the universal wavefunction for a weakly bound molecule in Eq.~\eqref{psiX-k}.
This failure of the derivation of their upper bound with $k_\mathrm{max}$ approximately equal to $\gamma_X$ 
has never been addressed by the authors of Ref.~\cite{Bignamini:2009sk}.

Assuming the binding energy of the $X$ is sufficiently small, we used the methods of Ref.~\cite{Artoisenet:2009wk}  to
derived the equality in Eq.~\eqref{sigmaX=sigmaDstarD}
between the $X$ cross section and the $D^{*0} \bar D^0$ cross section
integrated up to relative momentum $k_\mathrm{max}= 7.73\, \gamma_X$.
This equality takes into account the threshold enhancement in the charm-meson-pair cross section
associated with the $X$ resonance. 
The condition for the validity of this equality is that the binding energy $|E_X|$ is less than about 0.2~MeV.

The equality in Eq.~\eqref{sigmaX=sigmaDstarD} is not applicable if the charm-meson-pair cross section 
is estimated using a naive method that is not informed about the resonance, such as a Monte Carlo event generator.
We used the methods of Ref.~\cite{Artoisenet:2009wk}  to derive the order-of-magnitude estimate  
for the $X$ cross section in Eq.~\eqref{sigmaX=sigmaDstarDnaive}.
It is expressed as the naive $D^{*0} \bar D^0$ cross section
integrated up to a relative momentum $k_\mathrm{max}$ of order $(m_\pi^2  \gamma_X)^{1/3}$.
The resulting estimate for the prompt cross section for $X$ at the Tevatron 
using the naive $D^{*0} \bar D^0$ cross section in Eq.~\eqref{sigmaDD:Tevatron} 
is compatible with the experimental lower bound on the prompt $X$ cross section given in Eq.~\eqref{sigma:Tevatron}. 
Given the large uncertainty in $k_\mathrm{max}$ and the $k_\mathrm{max}^3$ scaling of the naive cross section,
there is little doubt that the corresponding estimate of the cross section at the LHC 
would be compatible with the measured value in Eq.~\eqref{sigma:LHC}.
We conclude that the prompt cross sections for $X$ at the Tevatron and at the LHC  
are compatible with the identification of $X(3872)$ as a weakly bound charm-meson molecule.


\begin{acknowledgments}
This work was supported in part by the Department of Energy under grant DE-SC0011726
and by the National Science Foundation under grant  PHY-1607190.
We thank T.~Skwarnicki and F.K.~Guo for valuable comments.
\end{acknowledgments}


\end{document}